%% file: text1.tex
\def\({ \left( }
\def\){ \right) }
\def\b{\begin{equation}}
\def\e{\end{equation}}
\def\={\ =\ }

\def\+{\ +\ }
\def\-{\ -\ }

\def\mumu{$\mu^+\mu^-$}
\def\ee{$e^+e^-$}
\def\rr{}

\documentstyle[epsfig]{aipproc}
\begin{document}
\title{MUON COLLIDER: INTRODUCTION AND STATUS}

\author{R. B.~Palmer for the Muon Collider Collaboration\thanks{Members of the Collaboration can be found at \textbf{http://www.cap.bnl.gov/mumu/}}}
\address{Physics Department\\
Brookhaven National Laboratory,\\
  Upton, NY 11973-5000, USA}
\maketitle

\begin{abstract}
Parameters are given of machines with center-of-mass (CoM) energies of 3 TeV
and 400 GeV but, besides a comment on
neutrino radiation, the paper concentrates on progress on the design of a
machine to operate at a light Higgs mass, assumed, for this study, to be
100 GeV (CoM).
\end{abstract}

\input{text2.tex}

\end{document}

%% file: text2.tex
\section*{INTRODUCTION}
\begin{figure}[bth!] 
\centerline{\epsfig{file=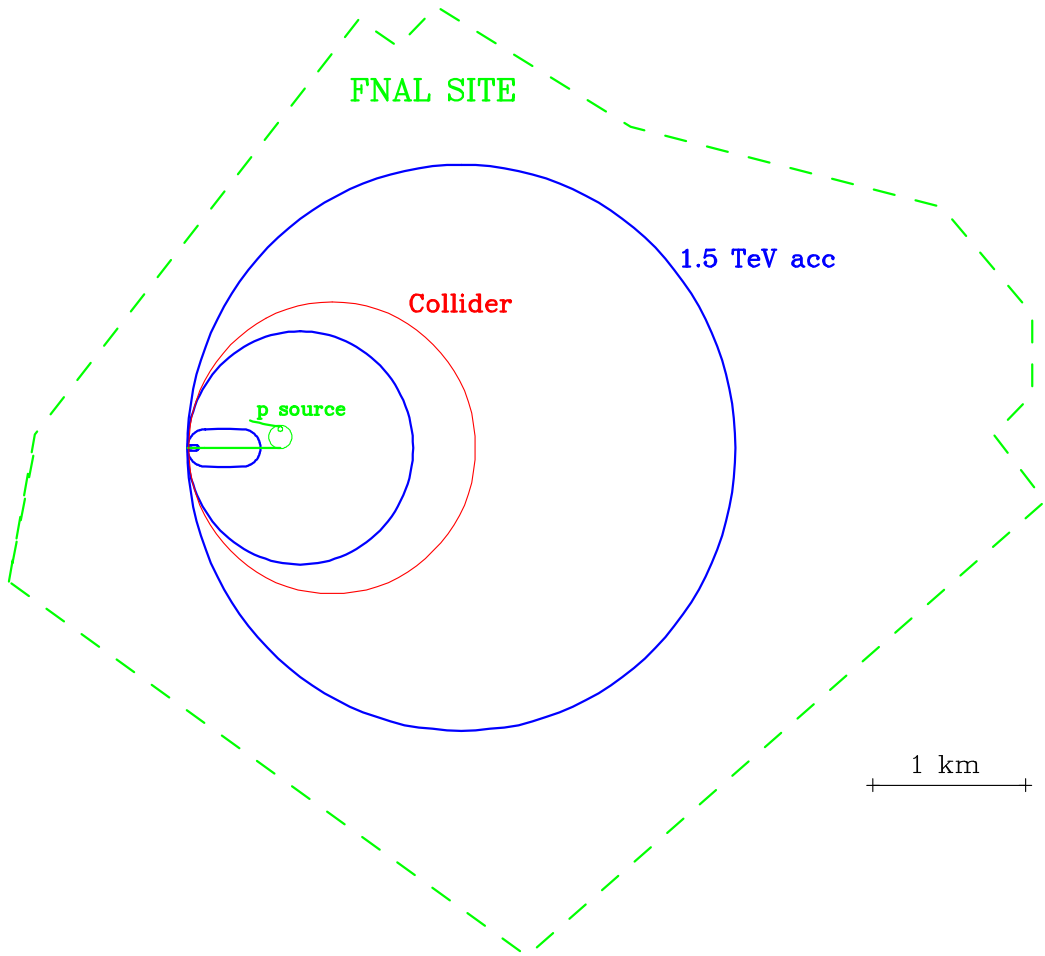,height=5.0in,width=5.45in}}
\caption{ Plan of a 3 TeV Muon Collider.}
\label{plan2}
\end{figure}

\begin{figure}[tbh!]
\centerline{\epsfig{file=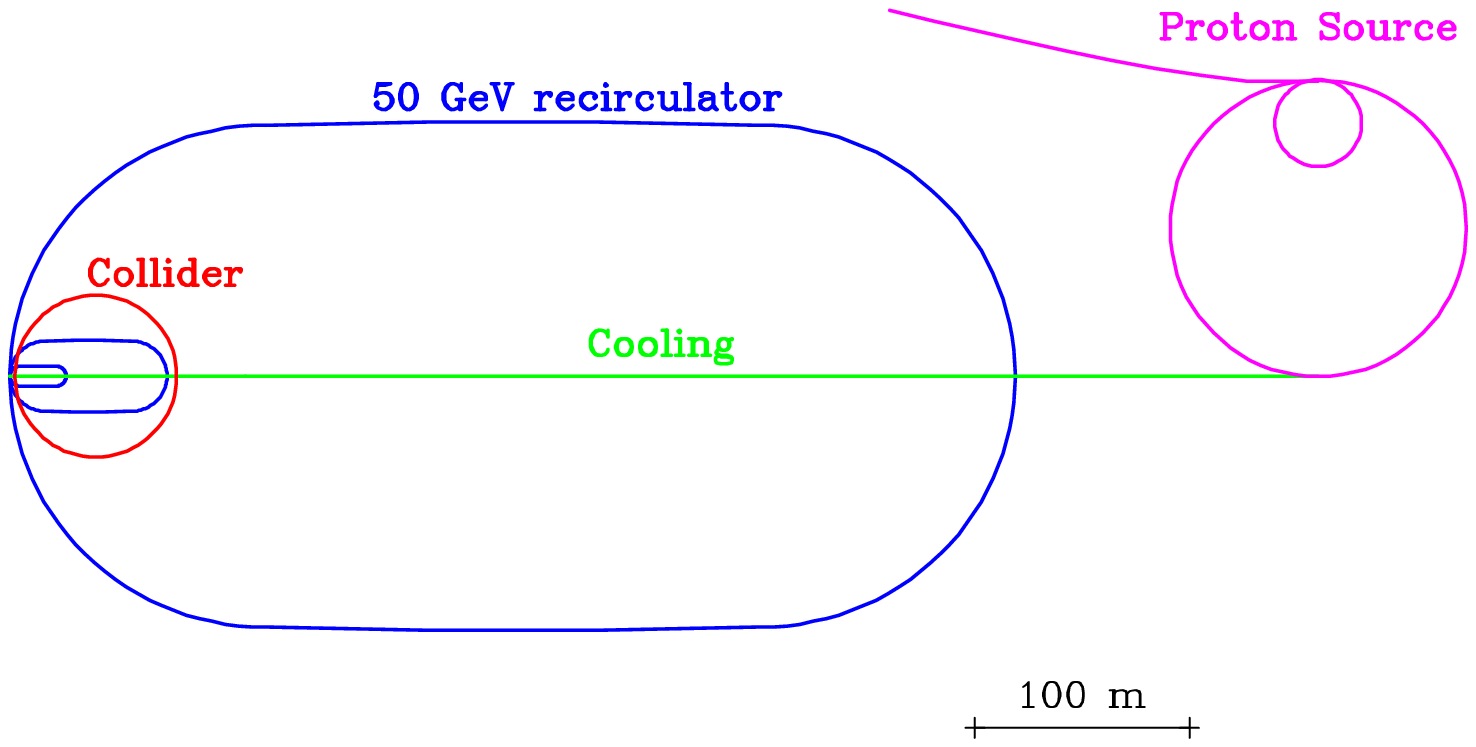,height=2.95in,width=5.75in}} 
\caption{ Plan of a 100 GeV Muon Collider.}
\label{plan1}
\end{figure}

The possibility of muon colliders was introduced  by Skrinsky et
al.\cite{ref1} and Neuffer\cite{ref2}. 
More recently, a collaboration of over 100 members, lead by BNL, FNAL, LBNL,
BNIP, University of Mississippi, Princeton University and UCLA has been formed to coordinate
studies on specific designs. Work has been done on designs at a 3-4 TeV,
0.4-0.5 TeV and 
$\approx$100 GeV\cite{ref3,ref4,ref5,ref6,ref6a}. 
Tb.~\ref{sum} gives the parameters of such colliders, and Figs.~\ref{plan2} and \ref{plan1}  show possible outlines of the 3 TeV and 100 GeV machines. 

\begin{table}[thb!]  
\caption{Parameters of Collider Rings}
\label{sum}
\begin{tabular}{|ll|c|c|ccc|}
\hline
\rr (CoM) energy         &\rr TeV   &\rr 3 &\rr 0.4 &
\multicolumn{3}{c|}{0.1 }  \\
\hline
p energy       & GeV        &  16  & 16 & \multicolumn{3}{c|}{16}\\
p's/bunch      & $10^{13}$    &  2.5 & 2.5 & \multicolumn{3}{c|}{5 }  \\  
bunches/fill   &           & 4 & 4 & \multicolumn{3}{c|}{2 }  \\
rep rate  & Hz     &  15 & 15 & \multicolumn{3}{c|}{15 }  \\
p power        & MW         &  4   & 4 & \multicolumn{3}{c|}{4}  \\ 
$\mu$/bunch  &$10^{12}$    & 2 & 2 &   \multicolumn{3}{c|}{4 }  \\
\rr $\mu$ power  &\rr MW     & \rr 28 &\rr 4 & \multicolumn{3}{c|}{\rr 1 }  \\
\rr wall power    &\rr MW    &  \rr  204 &\rr 120  & \multicolumn{3}{c|}{\rr
81 }  \\
collider circ   & m          &  6000 & 1000 & \multicolumn{3}{c|}{300 }  \\
depth   & m          &  500 & 100 & \multicolumn{3}{c|}{10 }  \\
\hline
\rr rms ${\Delta p\over p}$       &\rr  \%          &\rr .16 &\rr .14 &\rr .12 &\rr .01&\rr
.003 \\
\hline
6D $\epsilon_6$    & $10^{-12}\ (\pi m)^3$&170&170&170&170&170\\
rms $\epsilon_n$     &$\pi$ mm mrad     &  50 & 50 & 85 & 195 & 280\\
$\beta^*$         & cm          & 0.3 & 2.3 & 4 &  9 & 13\\
$\sigma_z$         & cm          & 0.3 & 2.3 & 4 &  9 & 13 \\
$\sigma_r$ spot    &$\mu m$     & 3.2 & 24 & 82 & 187 & 270\\
tune shift     &             &0.043 &0.043 & 0.05 &0.02 & .015\\
\hline
\rr Luminosity     &\rr $cm^{-2}sec^{-1}$&\rr 5 $10^{34}$ & $10^{33}$ &\rr
1.2 $10^{32}$ &\rr 2 $10^{31}$&\rr $10^{31}$ \\
(CoM) ${\Delta E\over E}$          & $10^{-5}$ & 80 & 80 & 80 & 7 & 2 \\ 
Higgs/year    & $10^3\ year^{-1}$ &  & & 1.6 & 4 & 4 \\
\end{tabular}
\end{table}

The original motive for considering muon colliders was the effective energy
advantage of any lepton collider over hadron machines, together with the
fact that muons, unlike electrons, generate negligible synchrotron
radiation. As a result, a muon collider can be circular and much smaller
than the current designs of linear electron colliders, and also
much smaller than a hadron machine with the same {\it effective} energy.

In addition, a \mumu collider would have some unique physics advantages over
an \ee collider:
   \begin{itemize}
\item The direct coupling of a lepton-lepton system to a Higgs boson has a
cross section that is proportional to the square of the mass of the lepton. As
a result, the cross section for direct Higgs production from the \mumu system
is 40,000 times that from an \ee system. 
\item Because of the lack of beamstrahlung, a \mumu collider can be operated
with an energy spread of as little as 0.003 \%. Furthermore, with the
naturally occurring polarization it would be possible, by observing g-2, to
determine the absolute energy to an accuracy of $10^{-6}$ or
better\cite{ref7}. It should thus be possible to use a
\mumu collider to make precision measurements of masses and direct
measurements of the Higgs width (assumed to be $\approx$ 2 MeV), that would be
otherwise impossible, with an \ee collider.
\end{itemize}

Machines with energies higher than 3-4 TeV, have significant beam current
constraints from off site neutrino radiation limits. If the required
luminosities are to be reached without unacceptable hazards,  then
significant improvements in muon emittance over the current base line values
are needed. There is however reason to believe such improvements are
achievable, and machines with a center of mass energy of 10 TeV and
luminosities of $10^{35}\ cm^{-2}s^{-1}$ and above may be
possible\cite{ref8}. For energies below 3 TeV, for fixed muon 
currents, this radiation falls as the energy cubed, and it should be little
problem for machines with energies of 1.5 TeV or less.

Recent work in the collaboration has concentrated on
the lowest energy machine ($\approx$ 100 GeV), whose energy is taken to be
representative of the possible mass of a light Higgs particle. Such a
machine would serve as a demonstration of Muon Collider technology, a needed
step before high energy machines can be considered, and as a unique physics
tool to make and study, if they exist, Higgs particles in the S-channel. 
\section*{PROTON DRIVER}
$\pi$ production rises approximately linearly with proton energy up to about 10 GeV after which it continues rising more slowly, but the requirement of very short bunches sets an effective minimum proton energy of about 16 GeV.
The baseline specification used in Tb.\ref{driver} is for a 16 GeV proton with a repetition rate of 15 Hz, $10^{14}$ protons per cycle in 2 or 4 bunches (depending on the collider energy), each with an rms bunch length of 1-2 ns. The total beam power is 4 MW.
   A design worked out at FNAL\cite{noble} would involve: a) An upgraded linac (0.4 $\rightarrow$ 1 GeV); b) higher energy booster (8 $\rightarrow$ 16 GeV) and c) new pre-booster. Some parameters are given in Tb.~\ref{driver}. 
\begin{table}[htb!] 
\caption{Proton Driver Specifications}
\label{driver}
\begin{tabular}{|ll|ccc|}
\hline
	     &     & Linac & Pre-Booster & Booster \\
\hline
Final energy & GeV & 1.0 & 4.5 & 16 \\
Protons/bunch &    &     & $5\ 10^{13}$ & $5\ 10^{13}$ \\
No of bunches &    &     &   2  &  2   \\
Rep. freq     & Hz &  15 &  15  &  15  \\
\hline
Circumference & m  &     &  180.6 & 474.2 \\
Norm. 95\% emit.&$\pi\ mm\ mrad$ & & 200 & 240 \\
sp ch tune shift    &    &     &   .39 &  .39  \\
Final field   & T  &     &   1.3   &  1.3 \\  
\end{tabular}

\end{table}

Another study had been done at BNL\cite{ref9} that, while it did not quite
reach the same beam power, involved far less upgrade: a) upgraded linac (0.2 $\rightarrow$ 0.6 GeV); b) increased AGS rep rate: 2.5 Hz.

 In order to reduce the cost
of the muon phase rotation section and for minimizing the final muon
longitudinal phase space, it appears now that the final proton bunch length
should be 1-2 ns. 

An experiment\cite{ref10} at the AGS has tested a method to generate such short bunches by rapidly bringing the tune of the machine near transition and allowing a strong phase rotation to occur. Bunches were shortened from 8 ns rms to 2.2 ns with initial 
longitudinal phase space similar to that specified in the above design. Shorter bunches are expected in later experiments with better control.

Another experiment\cite{lanl_inductor} has used variable inductors to reduce the longitudinal space charge effects.
\subsection*{Target and  Pion Capture}

$\pi$ production is maximized by the use a well focused proton beam, small diameter
target and a high Z target material. Tungsten, platinum or lead would be good, but the
heating could not be easily removed and shock damage could be a problem. The
use of a rapidly flowing liquid can solve the heating problem, but the shock
could damage the enclosure, if one is used. We are thus considering the use of
an open liquid jet. Such a jet has been tested\cite{ref11} using mercury,
although this was never exposed to a beam, and the jet did not move in a strong
magnetic field, as required in our case. Theoretical studies of liquid metal
flow in magnetic fields are underway\cite{ref12,ref13}, and the possibilities of using insulating
liquids (e.g. PtO$_2$, Re$_2$O$_3$) and slurries (e.g. Pt in water) are  being
considered.

If the axis of the target is coincident with that of the solenoid field, then
there is a relatively high probability that pions produced at the start of the
target will reenter, interact again later and be lost. The probability for such
interactions is reduced, and the overall production rate increased (by about 60
\%) if the target and proton beam are set at an angle (10-15$^o$) with respect to the field axis\cite{ref14,ref15}.

\begin{figure}[thb!] 
\centerline{\epsfig{file=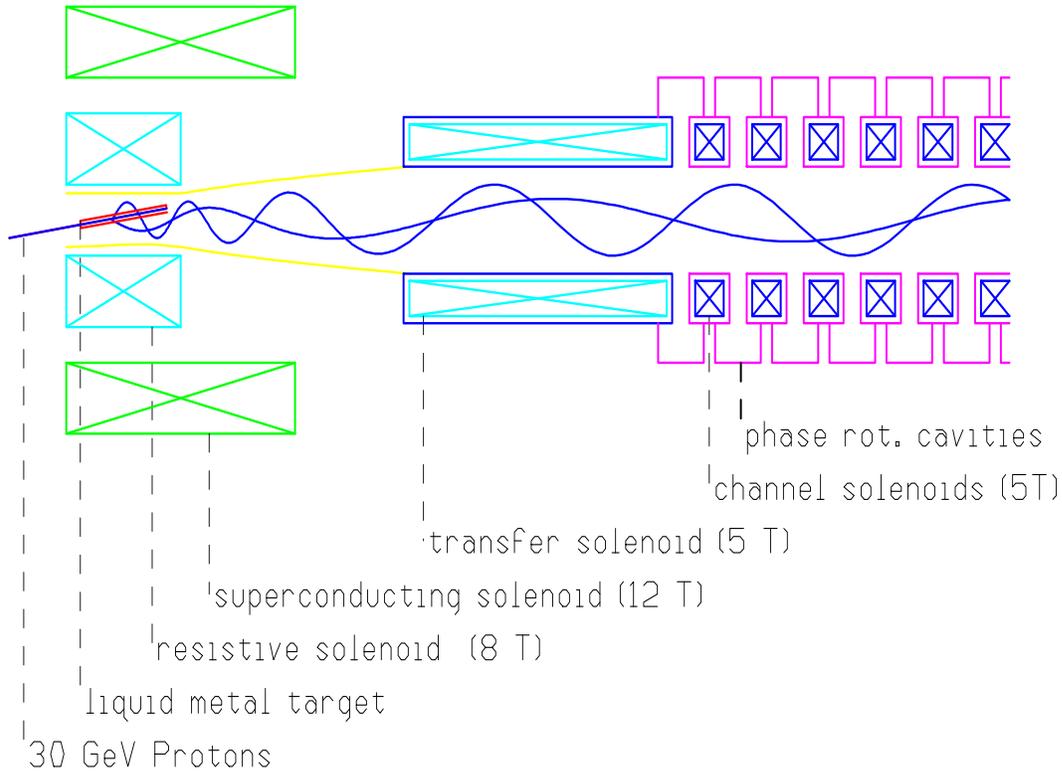,height=5.5in,width=4.in,angle=90}}
\vspace{0.5cm}
\caption{Schematics of the front end: skewed target, high field solenoid and decay and phase rotation channel }
 \label{capture}
 \end{figure}

Three different codes\cite{arc,mars,dpmjet} have been used to estimate $\pi$ yields and, despite detailed differences between them, overall $\mu$ production was very similar. In addition, the collaboration is involved in an AGS experiment\cite{exp910} to measure the $\pi$ yields. The production is peaked at a relatively low pion momentum 
($\approx$ 200 MeV/c), but has a very wide distribution: ${\Delta E\over E}$ rms $\approx$ 100
\%. The pion multiplicity, per 16 GeV proton, is about 2. At these low
energies, the transverse momenta are of the order of 200 MeV/c. If a
substantial fraction of these pions are to be captured, a very wide band system
is required. A 20 T solenoid, 16 cm inside diameter is found to capture about
half of all produced pions, and with target efficiency included, about 0.6
pions per proton emerge from the solenoid end\cite{ref16}. Such a
solenoid is well within the parameters of existing magnets\cite{ref17}. It
would have a superconducting outsert, and an 8  MW water cooled copper
insert\cite{ref18} (see Fig.~\ref{capture}).

After capture, the 20 T solenoid is matched\cite{snake} into a decay channel
with 5 T fields and diameter of 30 cm.
\subsection*{Phase Rotation Linac}
The pions, and the muons into which they decay, have an energy spread with an
rms value of approximately 100~\%. It would be difficult to handle such a wide
spread in any subsequent system. A linac is thus introduced along the decay
channel, with frequencies and phases chosen to deaccelerate the fast particles
and accelerate the slow ones; i.e. to phase rotate the muon bunch. Tb.\ref{rot}
gives an example of parameters of such a linac. It is seen that the lowest
frequency is 30 MHz, a low but not impossible frequency for a conventional
structure.
\begin{table}[bth!]
\caption{Parameters of Phase Rotation Linacs}
\label{rot}
\begin{tabular}{|c|c|c|c|}
\hline
Linac     & Length & Frequency & Gradient  \\
          &  m     &   MHz     &  MeV/m    \\
\hline
1         &  3    &   60     &   5      \\
2         &  29   &   30     &   4       \\
3         &  5    &   60      &   4      \\
4         &  5    &   37     &   4       \\
\end{tabular}
\end{table}

\begin{figure}[hbt!] 
\centerline{\epsfig{file=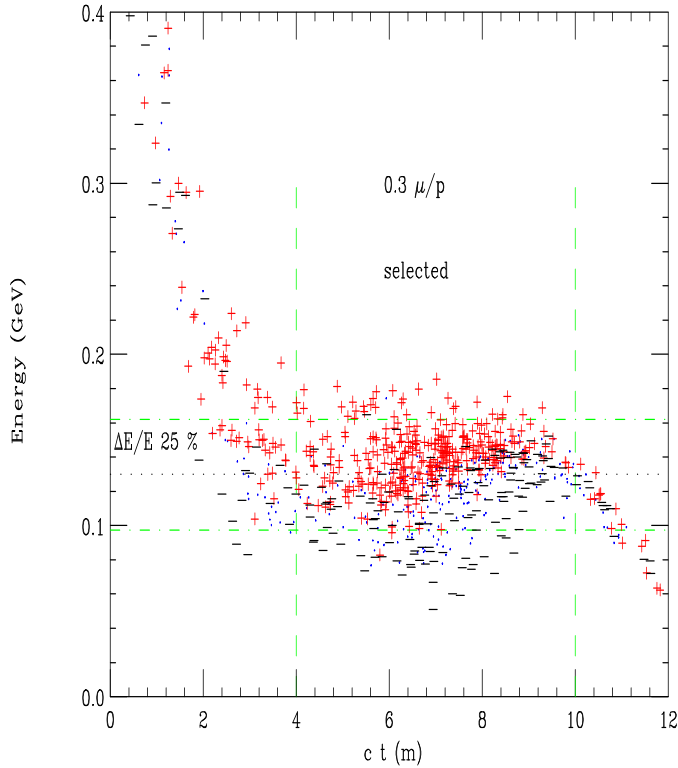,height=4.0in,width=3.5in}}
\caption{Energy vs. ct of $\mu$'s at end of decay channel with phase 
rotation.}
 \label{Evsctpol2}
 \end{figure}
 Fig.~\ref{Evsctpol2}  shows the energy vs. c$\,$t at the end 
of the decay and phase rotation channel. 
A bunch is defined with mean energy 150 MeV,
rms bunch length $1.7\,$m, and rms momentum  spread  $20\,$\% ($95\,$\%,
$\epsilon_{\rm L}= 3.2\,{\rm eV s}$) in the Monte Carlo study\cite{ref6}. The number of muons per initial proton in this selected  bunch was 0.38, which can be compared with a value of 0.3 assumed in the baseline parameters. 

\subsection*{ Use of Both Signs}
Protons on the target produce pions of both signs, and a solenoid will capture 
both, but the required subsequent phase rotation rf systems will have opposite 
effects on each. The baseline solution is to use two proton bunches, aim them 
at the same target one after the other, and adjust the rf phases such as to 
act correctly on one sign of the first bunch and on the other sign of the 
second. 

A second possibility would be to separate the charges into two channels, and
phase rotate them separately. However, the separation, probably using a bent
solenoid, is not simple and would not be fully efficient. Whether a gain in
overall efficiency could be achieved is not yet known.

\subsection*{Polarization}
\subsubsection*{Polarized Muon Production}

   In the center of mass of a decaying pion, the outgoing muon is fully 
polarized (-1 for $\mu^+$ and +1 for $\mu^-$). In the lab system the 
polarization depends\cite{ref19} on the decay angle $\theta_d$ and 
initial pion energy. For pion kinetic energy larger than the pion mass, the 
 average is about 20 \%, and if nothing else is done, the 
polarization of the captured muons and phase rotated by the proposed system is approximately this value.

\begin{figure}[bht!] 
\centerline{\epsfig{file=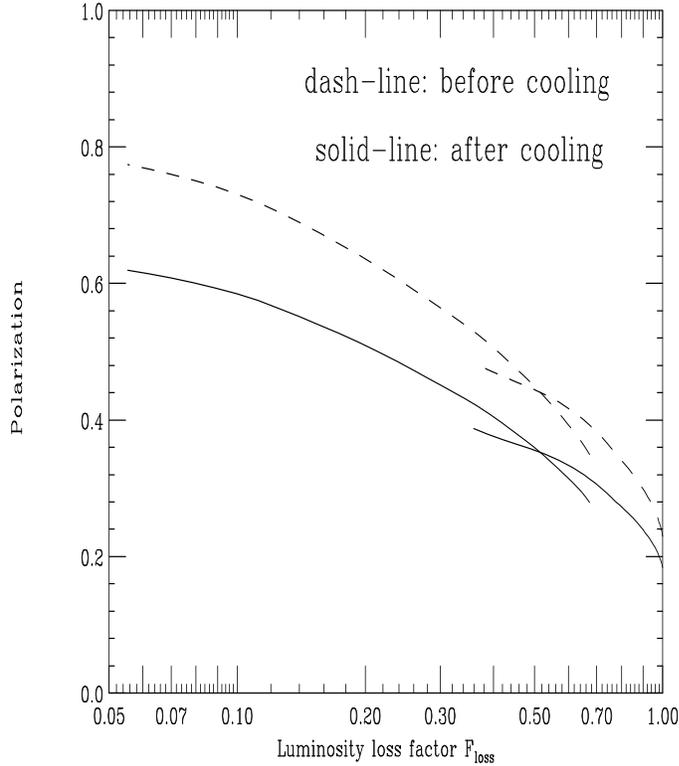,height=4.0in,width=3.5in}}
\caption{Polarization vs $F_{loss}$ of $\mu$'s accepted. 
\label{polvscutnew}}
\end{figure}

If higher polarization is required, some selection of muons from forward pion
decays  $(\cos{\theta_d} \rightarrow 1)$ is required. Fig.~\ref{Evsctpol2}, above, showed the polarization of the phase rotated muons. The polarization P$>{1\over 3}$,
$-{1\over 3}< P<{1\over 3}$, and P$<-{1\over 3}$ is marked by the symbols
$+,\,\bf{.}\,$ and $-$ respectively. 
If a selection is made on the minimum energy of the muons, then greater 
polarization is obtained. The tighter the cut, the higher the 
polarization, but the less the fraction $F_{loss}$ of muons that are selected. 
Fig.~\ref{polvscutnew} gives the results of a Monte Carlo study.

   If this selection is made on both beams, and if the proton bunch intensity is maintained, then naturally the muon bunch is reduced by the factor $F_{loss}$ and the luminosity would fall by $F_{loss}^2$. But if, instead, proton bunches 
are merged so as to obtain half as many bunches with twice the intensity, then the muon bunch intensity is maintained and the luminosity (and repetition rate)  falls only as $F_{loss}$.

   One also notes that the luminosity could be maintained at the full 
unpolarized value if the proton source intensity could be increased. Such an 
increase in proton source intensity in the unpolarized case would be 
impractical because of the resultant excessive high energy muon beam power, 
but this restriction does not apply if the increase is used to offset losses 
in generating polarization.

\subsubsection*{Polarization Preservation}

   A paper\cite{ref20} has discussed the preservation of muon
polarization in some detail. 
During the ionization cooling process the muons lose energy in material and 
have a spin flip probability ${\cal P},$ 
 \b
{\cal P}\approx \int {{m_e}\over {m_{\mu}}}\beta_v^2\ {\Delta E\over E }
 \e
where $\beta_v$ is the muon velocity divided by c, and ${\Delta E\over E}$ is the fractional 
loss of energy due to ionization loss. In our case the integrated energy loss 
is approximately 3 GeV and the typical energy is 150 MeV, so the integrated 
spin flip probability is close to 10~\%. The change in polarization 
${\Delta {\cal P}\over {\cal P}}$ is 
twice the spin flip probability, so the reduction in polarization is 
approximately 20~\%. 
This loss is included in Fig.~\ref{polvscutnew}.

   During circulation in any ring, the muon spins, if initially longitudinal, 
will precess by (g-2)/2 $\gamma$ turns per revolution; where
(g-2)/2 is $1.166\ 10^{-3}$. A given energy spread ${\Delta \gamma \over \gamma}$ will 
introduce variations in these precessions and cause dilution of the polarization. 
But if the particles remain in the ring for an exact integer number of synchrotron 
oscillations, then their individual average $\gamma$'s will be the same and no 
dilution will occur. 

   In the collider,
 bending can be performed with the spin orientation in the vertical 
direction, and the spin rotated into 
the longitudinal direction only for the interaction region. The design of such 
spin rotators appears relatively straightforward, but long. This might be a preferred solution at high energies but is not practical for instance, in the 100 GeV machine. An alternative is to use such a small energy spread, as in the Higgs factory, that though the polarization vector precesses, the beam polarization  does not become significantly diluted. 
\section*{COOLING}
For a collider, the phase-space volume must be reduced within a time of the order of the
$\mu$ lifetime. Cooling by synchrotron radiation, conventional stochastic
cooling and conventional electron cooling are all too slow. Optical stochastic
cooling\cite{ref21}, electron cooling in a plasma discharge\cite{ref22} and
cooling in a crystal lattice\cite{ref23} are being studied, but appear
difficult. Ionization cooling\cite{ref24} of muons seems relatively
straightforward.

\subsection*{Ionization Cooling Theory}

In ionization cooling, the beam loses both transverse and longitudinal momentum
as it passes through a material medium. Subsequently, the longitudinal
momentum can be restored by coherent reacceleration, leaving a net loss of
transverse momentum. 

The approximate equation for transverse cooling (with energies in GeV)  is:
  \begin{equation}
\frac{d\epsilon_n}{ds}\ =\ -\frac{dE_{\mu}}{ds}\ \frac{\epsilon_n}{E_{\mu}}\ +
\ \frac{\beta_{\perp} (0.014)^2}{2\ E_{\mu}m_{\mu}\ L_R},\label{eq1}
  \end{equation}
where $\epsilon_n$ is the normalized emittance, $\beta_{\perp}$ is the betatron
function at the absorber, $dE_{\mu}/ds$ is the energy loss, and $L_R$  is the
radiation length of the material.  The first term in this equation is the coherent
cooling term, and the second is the heating due to multiple scattering.
This heating term is minimized if $\beta_{\perp}$ is small (strong-focusing)
and $L_R$ is large (a low-Z absorber). 

The equation for energy spread  (longitudinal emittance) is:
 \begin{equation}
{\frac{d(\Delta E)^2}{ds}}\ =\
-2\ {\frac{d\left( {\frac{dE_\mu}{ds}} \right)} {dE_\mu}}
\ <(\Delta E_{\mu})^2 >\ +\
{\frac{d(\Delta E_{\mu})^2_{{\rm straggling}}}{ds}}\label{eq2}
 \end{equation}
where the first term is the cooling (or heating) due to energy loss, 
and the second term is the heating due to straggling.

Energy spread can be reduced by artificially increasing
${d(dE_\mu/ds)\over dE_{\mu}}$ by placing a transverse variation in absorber
density or thickness at a location where position is energy dependent, i.e. where there is
dispersion. The use of such wedges can reduce energy spread, but it
simultaneously increases transverse emittance in the direction of the
dispersion. Six dimensional phase space is not reduced.

\subsection*{Cooling Components}

We require a reduction of the normalized transverse emittance by almost three
orders of magnitude (from $1\times 10^{-2}$ to $5\times 10^{-5}\,$m-rad), and a
reduction of the longitudinal emittance by one order of magnitude.
This cooling is obtained in a series of cooling stages. In general, each stage
consists of two components:

 \begin{enumerate}
 \item a material in a strong focusing (low$\beta_\perp$) environment alternated with linac accelerators. These components will cool the transverse phase space.
 \item a lattice that generates dispersion, with absorbing material wedges introduced to interchange
longitudinal and transverse emittance.
 \end{enumerate}

Simulations have been performed on examples of each component using the program ICOOL\cite{ref25} which includes Vavilov distributions (with Landau and Gaussian limits) for dE/dx, and Moliere scattering distributions (with Rutherford limit). The only effects which are not yet included are space-charge and wake-field effects. Analytic vacuum calculations indicate that these effects should be significant, but not overwhelming. A correct simulation must be done before we are assured that no real problems exist. 

\subsubsection*{Transverse Cooling}

\begin{figure}[bht!]
\centerline{\epsfig{file=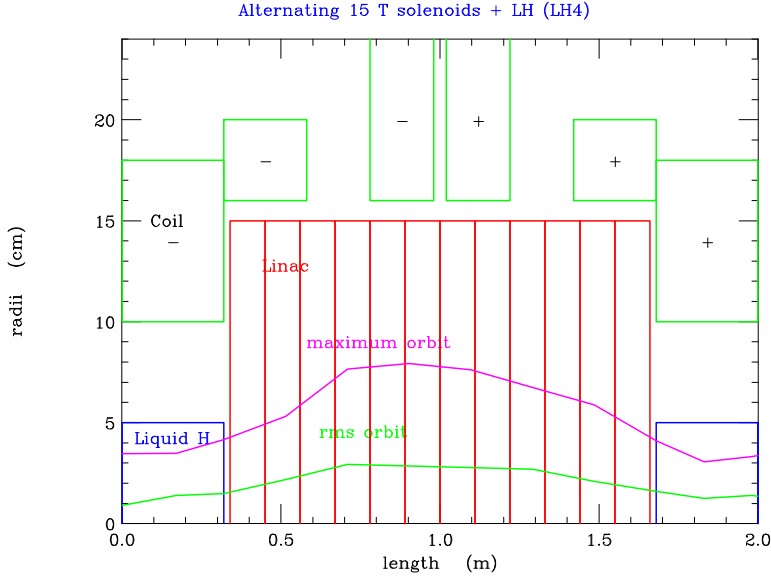,height=3.0in,width=4.0in}}
\caption{The cross section of one cell of an alternating solenoid cooling system }
 \label{altsol}
 \end{figure}

\begin{figure}[bht!]
\centerline{\epsfig{file=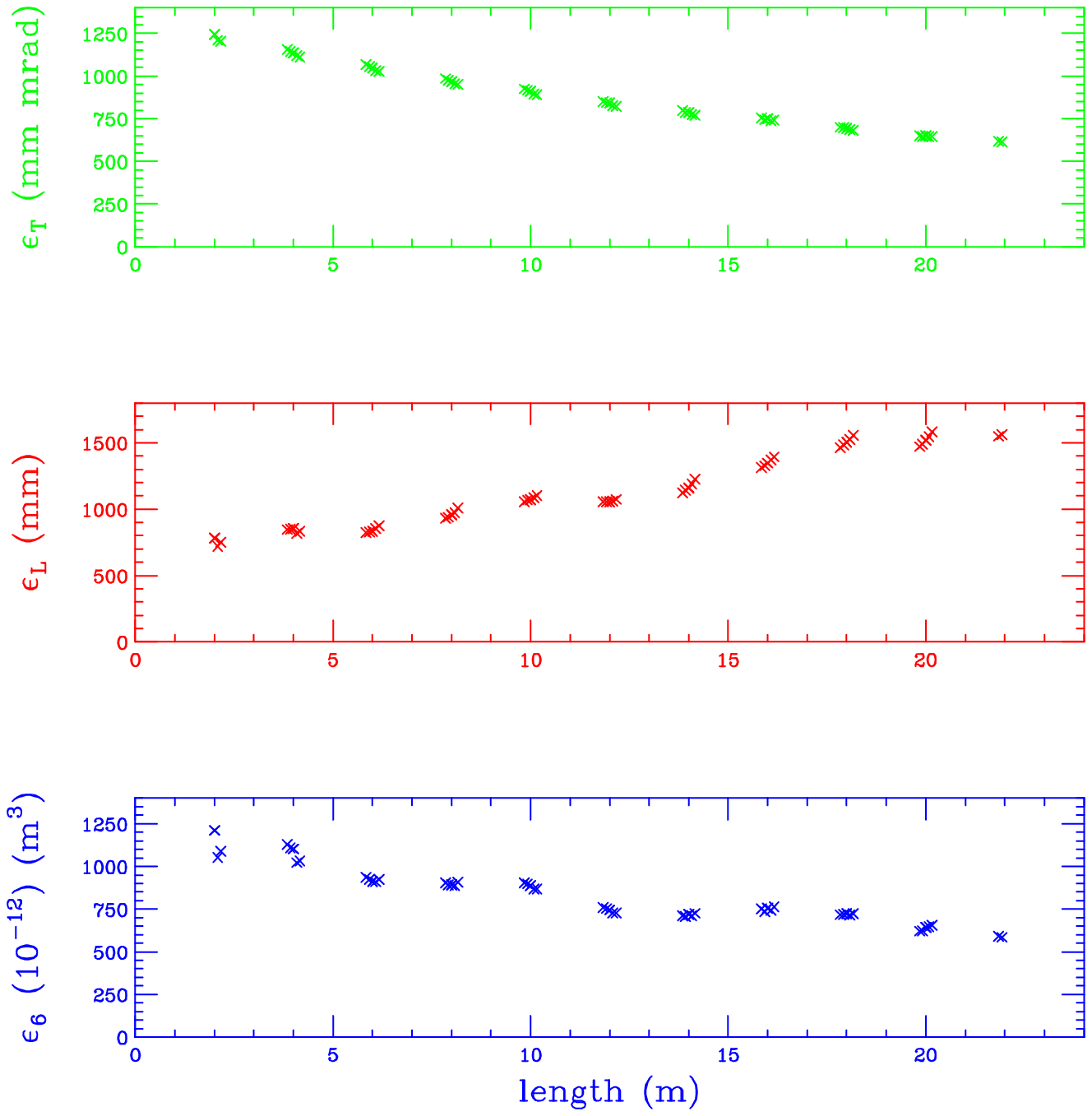,height=5.0in,width=5.0in}}
\caption{Emittance vs. length in 10 alternating solenoid cells; TOP: transverse emittance; MIDDLE: longitudinal emittance; and BOTTOM: 6-dimensional emittance}
 \label{coolingeg}
 \end{figure}

The baseline solution for the first component involves the use of liquid hydrogen absorbers in strong solenoid focusing fields,  interleaved with short linac sections. The solenoidal fields in succesive absorbers must be reversed to avoid build up of the canonical angular momentum. Fig.~\ref{altsol} shows the cross section of one cell of such a system. The top plot in Fig.~\ref{coolingeg} shows the reduction of transverse emittance in 10 such cells (20 m); the middle one shows the increase in longitudinal emittance induced by straggling and the adverse dependence of dE/dx with energy; while the bottom one shows the overall reduction in 6-dimensional emittance. This simulation has been confirmed, with minor differences by the codes double precision GEANT\cite{paul} and PARMELA\cite{parmela}.

Using 30 T solenoids at the end of a cooling sequence can attain a transverse emittance of 190 mm mrad and a six dimensional emittance of $30\times 10^{-12}$ m$^3$ (cf.280 mm mrad and  $170\times 10^{-12}$ m$^3$, respectively required for a Higgs factory). 

Other solutions, e.g. rapidly alternating solenoids and LiH absorbers\cite{fofo} and current carrying Li rods have been and will continue to be studied, but do not appear to be required to meet the baseline parameters (see below).
\subsubsection*{Linac}
The linacs used in the above simulations has a frequency of $805\,$MHz and required an accelerating gradient (peak phase) of $24\,$MV/m. The current designs use cavities separated by thin Be foils, ${2\pi \over 3}$ or ${2\pi \over 4}$ phase advanced per cavity, and powered in 3 of 4 separate interleaved side-coupled standing-wave systems\cite{zhao,moretti}. In order to reduce power source requirements the cavities may be operated at liquid nitrogen temperatures.

\subsubsection*{Longitudinal-Transverse Exchange}
\begin{figure}[bht!]
\centerline{\epsfig{file=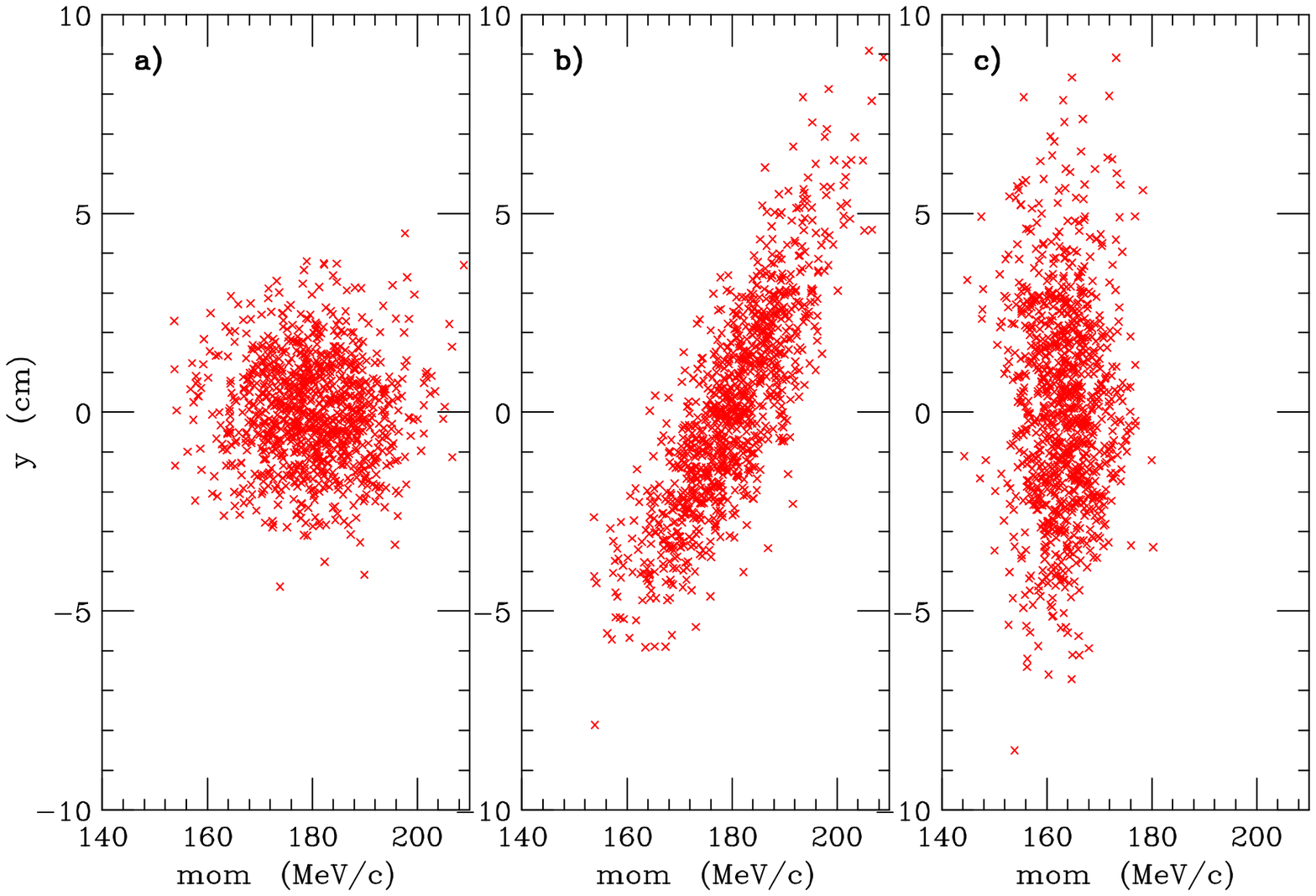,height=3.5in,width=5.5in}}
\vspace{0.5cm}
\caption{Transverse trajectory positions vs.their momenta: a) before the bend, b) after the bend, and c) after hydrogen wedges.}
 \label{exch}
 \end{figure}

The exchange of longitudinal and transverse emittance requires dispersion in a large acceptance channel. One way of achieving this is in a bent solenoid. Fig.~\ref{exch} shows transverse positions vs.their momenta: a) before the bend, b) after the bend, and c) after hydrogen wedges. The \textit{rms} momentum spread in this example is reduced from $8\,$MeV/c to $4.6\,$MeV/c with an accompanying approximate equivalent increase in the x-y emittance. 

Emittance exchange in solid wedges in the presence of ideal dispersion has also been simulated using SIMUCOOL\cite{dave}. Dispersion generation by weak focussing spectrometers\cite{balbekov} and dipoles with solenoids\cite{wan} are also studied.
\subsection*{Cooling System}
The required total 6 dimensional cooling is about 10$^6$. Since a single stage, as illustrated above, gives a factor of 2 reduction, about 20 such stages are required.
The total length of the system would be of the order of 500 m, and the total acceleration would be of the order of
6 GeV. The fraction of muons remaining at the end of the cooling system
is estimated to be $\approx 60 \%.$

In a few of the later stages, current carrying lithium rods might replace item
(1) above. In this case the rod serves simultaneously to maintain the low
$\beta_{\perp}$, and attenuate the beam momenta. Similar lithium rods, with
surface fields of $10\,$T , were developed at Novosibirsk (BINP) and have been used as focusing elements at FNAL and CERN\cite{ref26,ref27}. Cooling in beam recirculators could lead to reduction of costs of the cooling section\cite{balbekov}.
\section*{ACCELERATION}
Following cooling and initial bunch compression, the beams must be rapidly 
accelerated. A sequence of linacs would work, but would
be expensive, some form of circulating acceleration is preferred. At lower energies, the acceleration time is so short that any form of magnet ramping is probably impractical. The conservative option 
is to use
a sequence of recirculating accelerators (similar to that used at TJNL), but fixed frequency alternating gradient acceleration (FFAG) is also being studied\cite{ref28}. At higher energies, it is probably
 more economical to use fast rise time pulsed magnets in
more conventional synchrotrons\cite{ref29}.
\subsection*{Scenarios}
Tbs.~\ref{acceleration1} and \ref{acceleration2} give an example of possible sequences of
accelerators for a 100 GeV Higgs Factory and a 3 TeV collider. In both cases, following initial linacs, recirculating accelerators 
are used. Designs\cite{ref30} have been made of multiple aperture superconducting magnets for use in recirculating acceleration. The use of such magnets was not assumed in the scenarios, but they would reduce the diameter of the recirculating accelerator ring and lower particle loss from decay.

In the high energy case, the final three stages use pulsed magnet synchrotrons.
 If only pulsed magnets were used the power 
consumed by a ring would be high and its circumference large, but a hybrid ring with alternating 
pulsed warm magnets and fixed superconducting magnets appears practical. In the example, the last two such rings are located in the same tunnel with differing ratios of pulsed to fixed magnets. The fixed magnets are superconducting at 8 T; the pulse
d magnets are warm with fields that swing from -~2~T to +~2~T.

In both cases, except for the earliest stages, superconducting rf is employed. The reason for this, in the earlier stage, is that the instantaneous acceleration power requirement is very high, and the use of superconducting cavities allows a longer rf fill time and a  reduced rf power source requirement. For the higher energy accelerators the use of superconducting cavities is dictated by the need to achieve high wall to beam efficiency.

A study\cite{ref31} tracked particles through 
a similar sequence of recirculating accelerators and found a dilution of
longitudinal phase space of the order of 10\% and negligible particle loss.
\begin{table}[bth!]
 \caption{ Parameters Higgs Factory (100 GeV) Accelerators} 
\label{acceleration1}
\begin{tabular}{|ll|ccccc|c|}
\hline
acc type    &           &linac &linac &recirc&recirc&recirc& sums\\
rf type     &           &sledCu&sledCu&sledCu&sledCu&SC Nb &     \\
\hline 
$E_{init}$  &   (GeV)   &  0.10&  0.20&  0.70&    2 &    7 &      \\
\rr $E_{final}$ &\rr   (GeV)   &\rr  0.20&\rr  0.70&\rr    2 &\rr    7 &\rr   50 &      \\
\hline 
circ        &  (km)     &  0.04&  0.07&  0.06&  0.18&  1.21&  1.57\\
turns       &           &    1 &    1 &    8 &   10 &   11 &      \\
\rr decay loss  &\rr  (\%) &\rr  2.31&\rr  3.98&\rr  6.74&\rr  7.77&\rr 9.88&\rr 27.29\\
\rr decay heat  & \rr  W/m     &\rr  0.85&\rr  1.88&\rr 10.50&\rr 12.39&\rr 12.14&      \\
\hline 
$B_{fixed}$   &  (T)    &      &      &    2 &    2 &    2 &      \\
pipe width  &   cm      &      &      & 30.66& 21.22& 10.44&      \\
pipe ht     &   cm      &      &      &   10 &    8 &  4.30&      \\
\hline 
rf freq     &   (MHz)   &   90 &   90 &  120 &  170 &  400 &      \\
acc/turn    &   (GeV)   &  0.20&  0.40&  0.17&  0.50&    4 &      \\
acc time    & ($\mu s$) &      &      &    1 &    6 &   43 &      \\
acc Grad    &   (MV/m)  &    8 &    8 &    8 &   10 &   15 &      \\
grad sag    &    \%     &      &      & 13.08& 16.82& 27.15&      \\
rf time     &   ms    &  0.55&  0.56&  0.37&  0.24&  2.04&      \\
peak rf /m  &   (MW/m)  &  2.72&  2.56&  2.21&  4.40&  0.20&      \\
ave rf power&    MW     &  0.61&  1.10&  0.28&  0.88&  1.99&  4.88\\
\hline 
\rr total wall p& \rr  MW     & \rr 4.71&\rr  8.50&\rr  1.67&\rr  5.20&\rr  5.87&\rr 25.94\\
beam power  &    MW     &  0.00&  0.01&  0.03&  0.12&  0.92&  1.08\\
wall-beam eff&   \%     &  0.06&  0.15&  1.93&  2.22& 15.62&  4.16\\ 
\end{tabular}
\end{table}

\begin{table}[tbh!]
 \caption{ Parameters 3 TeV Collider Accelerators} 
\label{acceleration2}
\begin{tabular}{|ll|ccccccc|c|}
\hline
acc type    &           &linac &recirc&recirc&recirc&pulsed&pulsed &pulsed & sums\\
magnet type &           &      &warm  &warm  &warm  &warm  &hybrid&hybrid&     \\
rf type     &           &sledCu&sledCu&sledCu&SC Nb &SC Nb &SC Nb &SC Nb &     \\
\hline 
$E_{init}$  &   (GeV)   &  0.10&  0.70&    2 &    7 &   50 &  200 & 1000 &      \\
\rr $E_{final}$ &\rr (GeV) &\rr 0.70&\rr  2 &\rr 7 &\rr 50 &\rr 200 &\rr 1000 &\rr 1500 &      \\
\hline 
circ        &  (km)     &  0.07&  0.12&  0.25&  1.16&  4.65& 11.30& 11.36& 28.93\\
turns       &           &    2 &    8 &   10 &   11 &   15 &   27 &   17 &      \\
\rr decay loss  &\rr  (\%)     &\rr  6.11&\rr 12.11&\rr 10.38&\rr  9.53&\rr 10.68&\rr \rr 10.07&\rr  2.65&\rr 47.68\\
decay heat  &   W/m     &  3.46& 14.20& 16.03& 15.49& 19.44& 30.97& 18.09&      \\
\hline 
pulsed $B_{max}$   &    (T)    &      &      &      &      &    2 &    2 &    2 &      \\
$B_{fixed}$   &  (T)    &      &  0.70&  1.20&    2 &    2 &    8 &    8 &      \\
ramp freq   & (kHz)     &  900 &  109 & 40.02&  7.99&  1.43&  0.33&  0.53&      \\
\hline 
sig beam    &   cm      &  0.59&  0.51&  0.39&  0.25&  0.11&  0.08&  0.06&      \\
sig width   &   cm      &      &  3.03&  3.65&  3.85&  1.36&  0.59&  0.18&      \\
mom compactn&   \%      &      &   -1 &   -2 &   -2 &   -1 &   -1 &   -1 &      \\
pipe width  &   cm      &      & 30.31& 36.49& 38.53& 13.63&  5.86&    3 &      \\
pipe ht     &   cm      &      &   10 &    8 &  4.30&    3 &    3 &    3 &      \\
\hline 
rf freq     &   (MHz)   &   90 &   50 &   90 &  200 &  800 & 1300 & 1300 &      \\
acc/turn    &   (GeV)   &  0.40&  0.17&  0.50&    4 &   10 &   30 &   30 &      \\
acc time    & ($\mu s$) &      &    3 &    8 &   41 &  232 & 1004 &  631 &      \\
eta         &  (\%)     &  0.73&  0.22&  0.33&  0.44& 10.15& 14.37& 12.92&      \\
acc Grad    &   (MV/m)  &    8 &    8 &   10 &   15 &   15 &   25 &   25 &      \\
synch rot's &           &  0.54&  0.82&  1.91&  9.16& 27.07& 76.78& 31.30&      \\
phase slip  &  deg      &      &  6.90&  4.62&  5.35&  1.64&      &      &      \\
cavity rad  &   (cm)    &  122 &  220 &  134 & 76.52& 19.13& 11.77& 11.77&      \\
loading     &    \%     &  4.23&  6.22& 11.98& 16.54&  210 &  527 &  296 &      \\
grad sag    &    \%     &      &  3.16&  6.18&  8.65&      &      &      &      \\
rf time     &   msec    &  0.56&  1.35&  0.59&  2.04&  0.40&  1.25&  0.96&      \\
peak rf /m  &   (MW/m)  &  2.56&  3.43&  6.05&  0.81&  0.91&  0.56&  0.50&      \\
ave rf power&    MW     &  1.11&  1.54&  2.84&  7.20&  6.32& 21.91& 15.07& 55.99\\
\rr rf wall &\rr MW     &\rr  8.50&\rr  9.05&\rr 16.69&\rr 21.18&\rr 18.59&\rr 44.72&\rr 30.76&\rr  149 \\
\hline 
magnet ps   &   MJ    &      &      &      &      &      & 34.31& 13.19& 47.51\\
\rr magnet wall &\rr MW     &      &      &      &      &    &\rr 3.7&\rr 1.4& \rr 5.1\\
\hline 
\rr total wall  &\rr   MW &\rr  8.50&\rr  9.05&\rr 16.69&\rr 21.18&\rr 18.59&\rr 48.4& \rr 32.2&\rr  155 \\
beam power  &    MW     &  0.02&  0.04&  0.15&  1.17&  3.68& 17.54&  9.86& 32.47\\
wall-beam eff&   \%     &  0.26&  0.49&  0.91&  5.51& 19.81& 39.23& 32.06& 21.72\\
\hline 
\end{tabular}

\end{table}
\section*{COLLIDER STORAGE RING} 

After acceleration, the $\mu^+$ 
and $\mu^-$ bunches are injected into a separate storage ring. The highest 
possible average bending field is desirable, to maximize the number of 
revolutions before decay, and thus maximize the luminosity. Collisions would 
occur in one, or perhaps two, low-$\beta^*$ interaction areas. Parameters 
of the rings were given earlier in Tb.~\ref{sum}.
\subsection*{Lattice Design}
In order to 
maintain the required short bunches, without excessive rf, approximately 
isochronous  Flexible Momentum Compaction lattices\cite{ref32} would be used. 

In the high energy cases, the required betas at the intersection point are very small (e.g. $\beta^*=3\,{\rm mm}$ for 4 TeV), and the quadrupoles needed to generate them are large (20-30 cm diameter). At 100 GeV, the betas are not so small and the quadrupoles are more conventional, but in both cases it has been found that local chromatic correction is essential\cite{chromatic}.
\begin{figure}[bht!]
\centerline{\epsfig{file=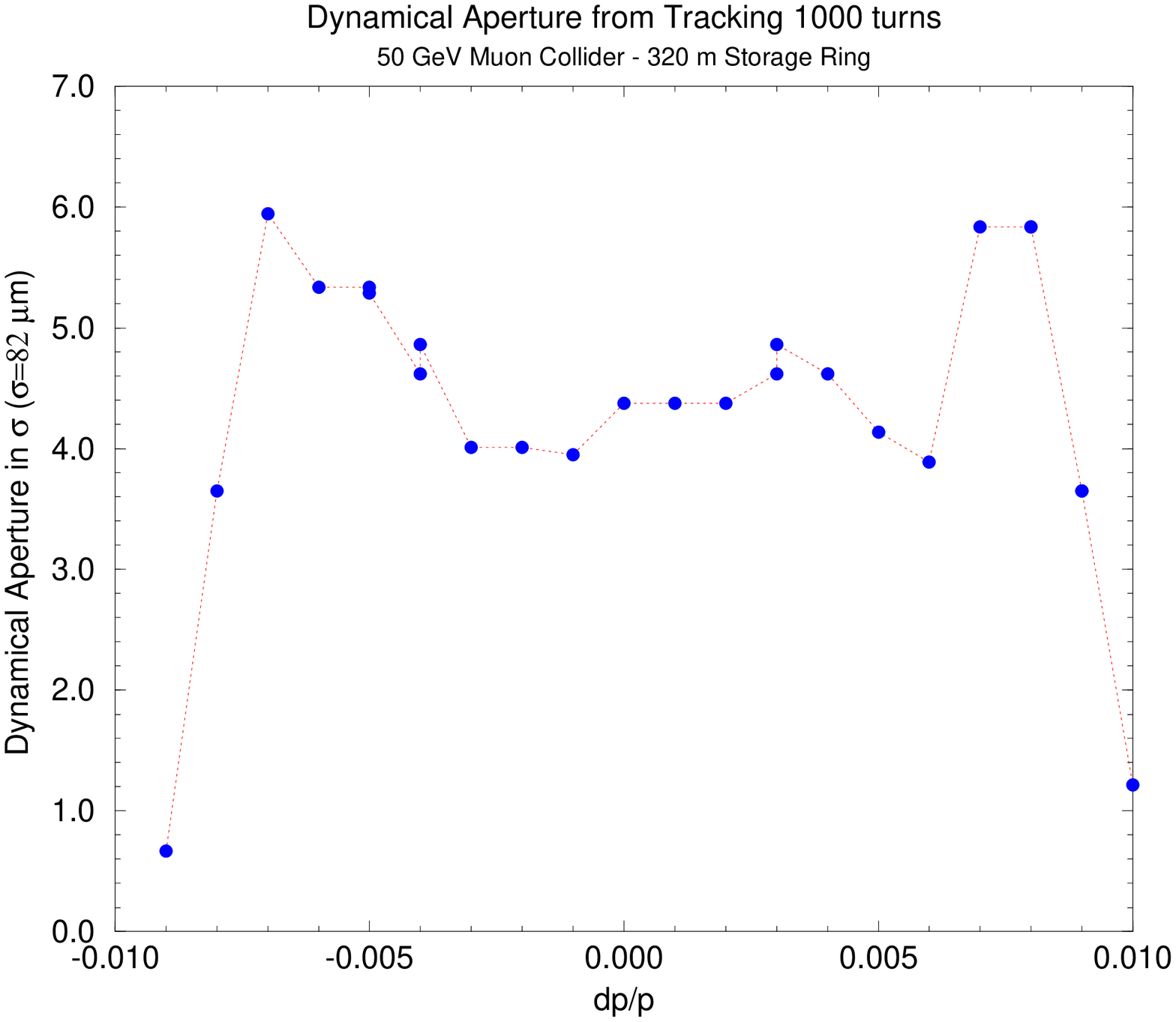,height=4.0in,width=3.5in, angle=-90}}
\caption{Dynamical aperture after 100 turns.}
 \label{dynaper}
 \end{figure}

Preliminary lattices have been designed for both 4 TeV and 0.5 TeV machines\cite{ref33}, and several designs now exist for the 100 GeV case. Fig.~\ref{dynaper} gives the dynamic aperture of one such lattice\cite{ref33a} for the required 1000 turns.
\subsection*{Scraping}
Collimation schemes have been designed\cite{ref34} for colliders at both high and low energies. At low energies, as in the Higgs Factory, tungsten collimators have been shown to be effective. At higher energies, the muons are scattered, but not stopped,
 by such collimators. For this case it has been shown that electrostatic septa followed by sweeping magnets could effectively extract the tail muons. Lattices\cite{ref33} have been designed incorporating these systems.
\subsection*{Instabilities}
The studies\cite{furman,chen} have considered beam emittance growth due to beam-beam tune shift, and both, althought some assumptions were made, predict negligible effects in 1000 tunes at the  values shown in Tb.~\ref{sum}.

A study\cite{ref35} has examined the resistive wall impedance longitudinal instabilities in rings at several energies. At the higher energies and larger momentum spreads, solutions were found with small but finite momentum compaction, and moderate rf. For the special case of the Higgs Factory, with its very low momentum spread, a solution was found with no synchrotron motion, but rf provided to correct the first order impedance generated momentum spread. The remaining off momentum tails, that would not affect the luminosity, but which might generate background, could be removed by a higher harmonic rf correction.

Given the very slow, or nonexistent synchrotron oscillations, the transverse beam breakup instability is significant. But this instability can
be  stabilized using rf quadrupole\cite{ref36} induced BNS damping. For instance, in the 3 TeV case, to stabilize the resistive wall instability, the required tune
spread, calculated\cite{ref38} using the two particle
model approximation, for a 1 cm radius aluminum pipe, is only 
$1.58\times10^{-4} $.

However, this application of the BNS damping to a quasi-isochronous ring, and other head-tail instabilities due to the  chromaticities $\xi$ and $
\eta_1$, needs more careful study. 
\subsection*{Bending Magnet Design}
\begin{figure}[bht!]
\centerline{\epsfig{file=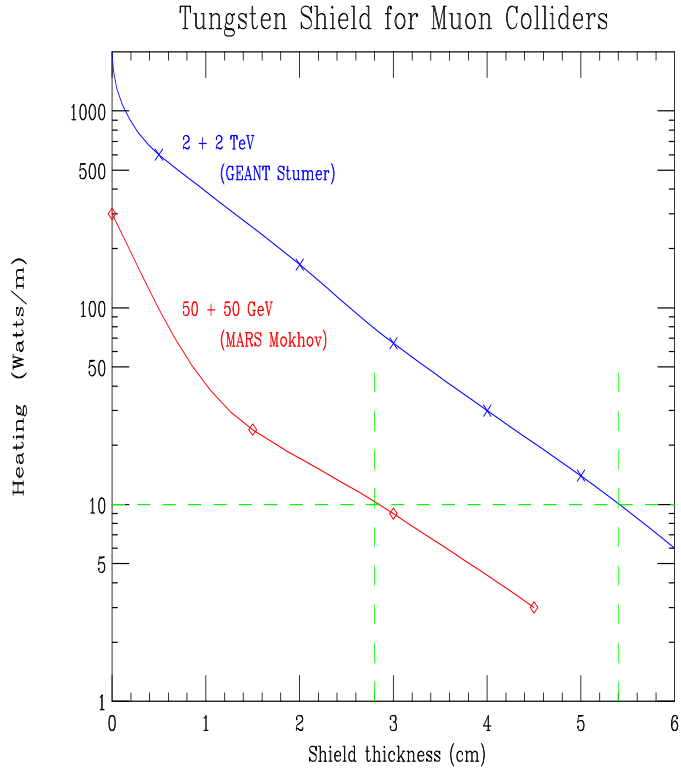,height=4.0in,width=3.5in}} 
\caption{Power penetrating tungsten shields vs. their thickness for, a) a 4 TeV, and b) a 100 GeV, Collider. \label{shieldingnew}}
 \end{figure}
The magnet design is complicated by the fact that the $\mu$'s decay within
the rings ($\mu^- \rightarrow\ e^-\overline{\nu_e}\nu_{\mu}$), producing
electrons whose mean energy is approximately $0.35$ that of the muons.   With no shielding, the average power deposited per unit length would be about 2 kW/m
in the 4 TeV machine, and 300 W/m in the 100 GeV Higgs factory.
Fig.~\ref{shieldingnew} shows the power penetrating tungsten shields of different thickness. One sees that 3 cm in the low energy case, or 6 cm at high energy would reduce the power to below 10 W/m which can reasonably be taken by superconducting magnets. 

The quadrupoles could use warm iron
poles placed as close to the beam as practical. The coils could then be either 
superconducting or warm, placed at a greater distance from the beam and shielded from it by the poles. 
\section*{NEUTRINO RADIATION}
Bruce King\cite{ref41} has shown that the surface radiation dose $D_B(Sv)$
in a time $t(s)$, in the plane of a bending magnet of field B(T), in
a circular collider with beam energy $E(TeV)$, average bending field
$<B(T)>$, at a depth $d(m)$ (assuming a spherical earth), with muon
current (of each sign) of $I\textrm{(muons/sec/sign)}$ is given by:
 $$
D_B\ \approx\ 4.4\ 10^{-24}\ {I_\mu \ E^3\ t \over d}\ {<B> \over B}\ t
 $$
and that the dose $D_S$ at a location on the surface, in line with a high
beta straight section of length $\ell$ is:
 $$
D_S\ \approx\ 6.7\ 10^{-24}\ {I_\mu \ E^3\ t \over d}\ {\ell  <B>}\ t
 $$
The first formula has been confirmed by a Monte Carlo simulation using
the MARS code\cite{mokhovrad}.
In all cases it is assumed that the average divergence angles satisfy the
condition: $\sigma_\theta << {1\over \gamma}.$
This condition is not satisfied in the straight sections approaching the IP,
and these regions, despite their length, do not contribute a significant dose.

  For the 3 TeV parameters given in Tb.~\ref{sum} (muon currents 
$I=6\times 10^{20}\,\mu^-/yr,$ $<B>= 6\, T,$ $B= 10\, T,$ and taking the federal limit on off site
radiation Dose/year, $D_{Fed}$ to be 1 mSv/year (100 mrem/year), then the
dose $D_B$ per year (defined as $10^7$ s), in the plane of a bending dipole
is:
 $$ 
D_B = 1.07 \ 10^{-5}\ \ (Sv)\ \approx 1 \% \ D_{Fed}
 $$
and for a straight section of length 0.6 m is:
 $$ 
D_B = 9.7 \ 10^{-5}\ \ (Sv)\ \approx 10 \% \ D_{Fed}
 $$
which may be taken to be a reasonable limit.

  Special care will be required in the lattice design to assure that no
field free region longer than this is present. But it may be noted that the
presence of a field of even 1 T over any length, is enough to reduce the dose to the 10
\% Federal limit standard. For machines above 3 TeV, the muon current would probably have to be reduced.

  For lower energy machines, the requirements get rapidly easier: a 0.5
TeV at 500 m depth could have 130 m straight sections, or if at 100 m depth
25 m lengths, for the same surface dose. For a 100 GeV machine the doses are
negligible. 
\section*{Detector Background}
There will be backgrounds in the detector from the decay of muons in the ring and approaching the IP. A recent study\cite{ref39} of electromagnetic, hadronic and muon components of the background has been done using the GEANT codes\cite{ref40}. 
This study
   \begin{itemize}
   \item
followed shower neutrons and photons down to 
$40\,$keV and electrons to $25\,$KeV.
   \item 
Used a tungsten shield over the beam, extending in to within 14 cm of the intersection point,
and extending outward to an angle of
20 degrees from the axis.
   \item
Inside this shield, between its smallest aperture 1 m from the IP, and its tip, the inner surface is shaped into a series of rising collimating steps and slopes, designed so that, the detector could not {\it see} any 
surface directly illuminated by the initial decay electrons, whether seen in the 
forward or backward (albedo) directions.                       
     \item
From the aperture point of minimum  to a few meters (2.5 m for Higgs Factory) upstream, the inside forms another series of stepped collimators placed at $\pm\ 4\ \ \sigma_{\theta_0}$ 
(where $\sigma_{\theta_0}$ is the rms divergence of the beam).
     \item
Further upstream, prior to the first quadrupole (from 2.5 to 4 m in the Higgs case) an 8 T dipole, with collimators inside, is used to sweep decay electrons before the final collimation.
   \end{itemize}

  Tb.~\ref{background} gives the hit density for the Higgs factory from the various sources and the occupancy of pixels of the given sizes.
 In all cases the numbers are given per bunch crossing. The hit density for the higher energy machines are found to be somewhat lower than these, due to the small decay angles of the electrons.
\begin{table}\caption{Detector backgrounds from $\mu$ decay}
\label{background}
\begin{tabular}{|llcccc|}
\hline
Radius             & cm   &      5   & 10  &20 & 100   \\
\hline
Photons hits     &$cm^{-2}$& 26 & 6.6 & 1.6 & .06   \\
Neutrons hits    &$cm^{-2}$ & 0.06 & 0.08 & 0.2 & 0.04     \\
Charged hits     &$cm^{-2}$ &8 & 1.2 & 0.2 & 0.01    \\
Total hits       &$cm^{-2}$ & 34 & 8 & 2 & 0.12   \\
\hline
Pixel size     & $\mu m^2$& 60x150&60x150&300x300&300x300 \\
Occupancy all    & \%  & 0.6 & 0.14 & 0.4 & 0.02 \\
Occupancy charged & \% & 0.14& 0.02 & 0.04 & 0.002 \\
\end{tabular}

\end{table}
The radiation damage by the neutrons on a silicon detector has also been
estimated. In the Higgs case, at 5 cm from the vertex, the number of hits from
neutrons above 100 KeV is found to be $1.8\, 10^{13}$ per year ($10^7\,s$). This is an order of magnitude less than that  expected at the LHC. The damages for silicon detectors in the higher energy machines are of the same order.

This study also found a significant flux of muons, with quite high energies,
from $\mu$ pair production in electromagnetic showers (Bethe Heitler). Their
most serious effect appears to arise when they make deeply inelastic
interactions and deposit spikes of energy in the electromagnetic and
hadronic calorimeters. This is not serious in the Higgs case, when the fluxes
and cross sections are low, but at higher energies timing and/or longitudinal
calorimeter segmentation appears necessary to identify and remove the problem.

An ealier study using the code MARS\cite{ref42}, using less sophisticated shielding, gave results qualitatively in agreement with those from Geant\cite{ref39}.

\section*{CONCLUSION}
\subsection*{Motive}
 \begin{itemize}
 \item Because they can be circular, muon colliders appear to be far smaller than hadron or \ee machines of similar effective energy.
 \item It is thus hoped that muon colliders could have a lower cost per TeV than other options.
 \item Their smaller size would allow machines up to 3 TeV effective energy (roughly equivalent to a 30 TeV hadron machine) to fit on existing laboratory sites.
 \item The low synchrotron and beamstrahlung radiation with muons could allow energy spreads as small as 0.003 \% (3 x 10$^{-5}$).
 \item By measuring g-2 of the muon it should be possible to determine the energy to even greater precision.
 \item The above, plus the large cross section for s-channel production, could make a muon collider into a precision tool to study Higgs particles and their decays.
 \end{itemize}

\subsection*{Progress}
 \begin{itemize}
 \item
The theory of operation of all components of a muon collider are now well understood.
 \item
Simulations of examples of all components of a baseline design have now been performed. The simulated performances of many of these components has exceeded the baseline specifications. All known effects have been included except space-charge in the cooling, whose effect, calculated analytically, appears not to be too large.

\subsection*{Needed}
 \item
More detailed simulations of all components, including space-charge in the cooling.
 \item
Complete scenarios of the cooling stages and acceleration.
 \item
 An experimental study of the target.
 \item
The construction and test of one or more of the cooling stages.
 \item 
Technical development of components: a large high field
solenoid for capture, low frequency rf linacs, multi-beam or
pulsed magnets for acceleration, warm bore shielded high field dipoles for the collider, muon collimators and background shields,
etc.. But none of these components can be described as {\it exotic}, 
and their specifications are not beyond what has been demonstrated.
 \end{itemize}

\section*{ACKNOWLEDGMENTS}

\medskip
This research was supported by the U.S. Department of Energy under Contract No.
DE-ACO2-76-CH00016 and DE-AC03-76SF00515.

%% file: text1.bbl
\begin{thebibliography}{99}

\bibitem{ref1}V. V. Parkhomchuk and A. N.
Skrinsky, Proc. 12th Int. Conf. on High Energy 
Accelerators, F. T. Cole and R.
Donaldson, Eds., (1983) 485; A. N. Skrinsky and V.V. 
Parkhomchuk, Sov. J. of
Nucl. Physics {\bf 12}, (1981) 223; {\it Early Concepts 
for $\mu^+\mu^-$
Colliders and High Energy $\mu$ Storage Rings}, {\it 
Physics Potential \&
Development of $\mu^+\mu^-$ Colliders. 2$^{nd}$ 
Workshop}, Sausalito, CA, Ed.
D. Cline, AIP Press, Woodbury, New York, (1995).

 \bibitem{ref2}D. Neuffer, Fermilab Note FN-319, July 1979; Proc. 12th Int. Conf. on High Energy Physics (1983) 481; {\it Principles and Applications of Muon Cooling}, Part. Acc. {\bf 14} 75 (1983) 

\bibitem{ref3}{\it Proceedings of the Mini-Workshop on 
$\mu^+\mu^-$  Colliders:
Particle Physics and Design}, Napa CA, Nucl Inst. and 
Meth., {\bf A350} (1994)
; Proceedings of the Muon Collider Workshop, February 
22, 1993, Los Alamos
National Laboratory Report LA-UR-93-866 (1993) and {\it 
Physics Potential \&
Development of $\mu^+\mu^-$ Colliders 2$^{nd}$ 
Workshop}, Sausalito, CA, Ed. D.
Cline, AIP Press, Woodbury, New York, (1995); Proceedings of
the 9th Advanced ICFA Beam Dynamics Workshop, Ed. J. C. 
Gallardo, AIP Press, Conference Proceedings 372 (1996).

\bibitem{ref4}R. B. Palmer et al., {\it Monte Carlo 
Simulations of Muon
Production}, {\it Physics Potential \& Development of 
$\mu^+\mu^-$ Colliders
2$^{nd}$ Workshop}, Sausalito, CA, Ed. D. Cline, AIP 
Press, Woodbury, New York,
pp. 108  (1995); R. B. Palmer, et al., {\it Muon 
Collider Design}, in Proceedings of the Symposium on 
Physics Potential \& Development of $\mu^+\mu^-$ 
Colliders, Nucl. Phys B (Proc. Suppl.) {\bf 51A} (1996); 
R. B. Palmer and J. C. Gallardo, {\it Muon-Muon and other High Energy Colliders}, \textit{Techniques and Concepts of High Energy Physics IX}, Phys. vol. 365,  Ed. T. Ferbel, pp. 183, Plenum Pub. (1997).

\bibitem{ref5}R. B. Palmer and J. C. Gallardo, \textit{ High Energy Colliders}, Proceedings of 250$^{th}$ Anniversary Conference on Critical Problems in Physics, Princeton University, Ed. Fitch, Marlow, Dementi, pp. 247 (1997). 

\bibitem{ref6}R. B. Palmer, \textit{ Progress on $\mu^+\mu^-$ Colliders }, submitted to the Proceedings of the PAC97, Vancouver, Canada, June 1997.

\bibitem{ref6a}{\it $\mu^+\mu^-$ Collider, A Feasibility Study},
BNL-52503, FermiLab-Conf-96/092, LBNL-38946, Proceedings of the 1996 DPF/DPB Summer Study on High-Energy Physics, Snowmass'96. For updated information, see the Muon Collider Collaboration WEB page: http://www.cap.bnl.gov/mumu/.

\bibitem{ref7}R. Raja and A. Tollestrup, \textit{Calibrating the energy of a 50 x 50 GeV muon collider using spin precession}, LANL preprint archive, hep-ex/9801004; submitted to Phys. Rev. D.
\bibitem{ref8}B. King, private communication
\bibitem{noble}C. Ankenbrandt and B. Noble, \textit{Summary of the Accelerator Working Group}, submitted to the Proceedings of Workshop on Physics at the First Muon Collider and at the Fron End, FNAL, Nov. 1997.
\bibitem{ref9}T. Roser, {\it AGS Performance and Upgrades: A Possible Proton
Driver for a Muon Collider}, Proceedings of
the 9th Advanced ICFA Beam Dynamics Workshop, Ed. J. C. 
Gallardo, AIP Press, Conference Proceedings 372 (1996).
\bibitem{ref10}C. Ankenbrandt, K-Y. Ng, J. Norem, M. Popovic, Z. Qian, L. Ahrens, M. Brennan, V. Mane, 
T. Roser, D. Trbojevic, W. van Asselt, \textit{ Bunching Near Transition in the AGS}, Fermilab Pub-98-006, 
submitted to Phys. Rev. D.
\bibitem{lanl_inductor}J. E. Griffin, K.Y. Ng, Z.B. Qian and D. Wildman, \textit{Experimental Study of Passive Compensation of Space Charge Potential Well Distortion at the Los Alamos National Laboratory Proton Storage Ring}, Fermilab Report, FN-661, Nov. 1997.

\bibitem{ref11}C. Johnson, \textit{Solid and Liquid Targets Overview}, presentation at the Mini-Wokshop: Target and Muon Collection Magnets and Accelerators, Oxford, MI, Jan. 1997. 
\bibitem{ref12}C. Lu, K. T. McDonald, \textit{Low-Melting-Temperature Metals for possible Use as Primary Targets at a Muon Collider Source}, Princeton/$\mu \mu$/97-3, Revised Dec. 1997, unpublished. 
\bibitem{ref13}R. Weggel, \textit{Deceleration of Conductor by Magnetic Field: 1) Paraxial; 2) Perpendicular}, unpublished
\bibitem{ref14}M. Green and R. Palmer, \textit{A $\mu-\mu$ collider capture solenoid system for pions froma tilted target}, submitted to the Proceedings of PAC97, May 1997. 
\bibitem{ref15}N.V. Mokhov and A. Van Ginneken, \textit{Pion Production and Targetry
at $\mu^+\mu^-$ Colliders}, Fermilab-Conf-98/041 (1998), submitted to
Proc. of the 4th Int. Conf. on Physics Potential and Development
at $\mu^+\mu^-$ Colliders, San Francisco, CA, December 10-12, 1997
   \bibitem{arc} D. Kahana, et al., {\it Proceedings of Heavy Ion Physics at the
AGS-HIPAGS '93}, Ed. G. S. Stephans, S. G. Steadman and W. E. Kehoe (1993); D.
Kahana and Y. Torun, {\it Analysis of Pion Production Data from E-802 at 14.6
GeV/c using ARC}, BNL Report \# 61983 (1995).
   \bibitem{mars} N. V. Mokhov, {\it The MARS Code System User's Guide},
version 13(95), Fermilab-FN-628 (1995).
\bibitem{dpmjet}J. Ranft, DPMJET Code System (1995).
\bibitem{exp910}Experiment E-910 at AGS, BNL, private communication.
\bibitem{ref16}N.V. Mokhov and S.I. Striganov, \textit{Towards Reliable Prediction of 
Particle Production for 6-120 GeV Proton Beams}, presentation at the Workshop on Physics at the First Muon Collider and at the Front End of a Muon Collider, Nov. 1997.
\bibitem{ref17}J. R. Miller, M. Bird, S. Bole et al., \textit{An Overview of the 45 T Hybrid Magnet System for the National High Field Magnet Laboratory}, IEEE Transactions on Magnetics 30, pp. 1563 (1994).
\bibitem{ref18}R. Weggel, \textit{4-MW Hollow-Conductor Magnets for 20 T Hybrid Systems to Collect Pions for a Muon Collider}, presentation at the Mini-Wokshop: Target and Muon Collection Magnets and Accelerators, Oxford, MI, Jan. 1997.
   \bibitem{snake}N. Mokhov, R. Noble and A. Van Ginneken, {\it Target and
Collection Optimization for Muon Colliders}, Proceedings of
the 9th Advanced ICFA Beam Dynamics Workshop, Ed. J. C. 
Gallardo, AIP Press, Conference Proceedings 372 (1996).  
\bibitem{ref19}K. Assamagan, et al., Phys. Lett. {\bf B}335, 231 (1994); E. P.
Wigner, Ann. Math. {\bf 40}, 194 (1939) and Rev. Mod. Phys., {\bf 29}, 255 (1957).
\bibitem{ref20}B. Norum and R. Rossmanith, {\it Polarized Beams in a Muon
Collider}, in Proceedings of the Symposium on 
Physics Potential \& Development of $\mu^+\mu^-$ 
Colliders, Nucl. Phys B (Proc. Suppl.) {\bf 51A} (1996).
\bibitem{ref21}A. A. Mikhailichenko and M. S. Zolotorev, Phys. Rev. Lett. {\bf
71}, (1993) 4146; M. S. Zolotorev and A. A. Zholents, SLAC-PUB-6476 (1994).
 \bibitem{ref22}A. Hershcovitch, Brookhaven National Report AGS/AD/Tech. Note
No. 413 (1995).
 \bibitem{ref23}
Z. Huang, P. Chen and R. Ruth, SLAC-PUB-6745, {\it Proc. Workshop on Advanced
Accelerator Concepts}, Lake Geneva, WI , June (1994); P. Sandler, A. Bogacz and
D. Cline, {\it Muon Cooling and Acceleration Experiment Using Muon Sources at
Triumf},  Physics Potential \& Development of $\mu^+\mu^-$ Colliders 2$^{nd}$
Workshop, Sausalito, CA, Ed. D. Cline, AIP Press, Woodbury, New York, pp. 146 
(1995).
 \bibitem{ref24}Initial speculations on ionization cooling have been variously
attributed to G. O'Neill and/or G. Budker see D. Neuffer in \cite{ref3};  D.
Neuffer, in Advanced Accelerator Concepts, AIP Conf. Proc. 156, 201 (1987); see
also \cite{ref1,ref2,ref3}; R. C. Fernow and J. C. Gallardo, {\it Muon Transverse Ionization Cooling: Stochastic Approach}, Phys. Rev. {\bf E52} 1039 (1995).
\bibitem{ref25}R. Fernow, \textbf{ICOOL}, fortran program to simulate muon ionozation cooling.
\bibitem{paul}P. Le Brun, \textit{Alternate solenoid in DPGeant}, presented at the Mini-Workshop on Cooling, BNL Jan. 1998, unpublished.
\bibitem{parmela}H. Kirk, \textit{Parmela modeling of alternating solenoids} presented at the Mini-Workshop on Cooling, BNL Jan. 1998, unpublished.
\bibitem{fofo}R. C. Fernow, J. C. Gallardo and R. B. Palmer, \textit{Ionization cooling using a FOFO lattice}, BNL Report BNL \#64493, submitted to PAC97, Vancouver, Canada, 1997.
\bibitem{zhao}Y. Zhao, \textit{The preliminary simulation of ${2\pi \over 3}$-mode interleaved side coupled standing wave structures}, presented at the Mini-Workshop on Cooling, BNL Jan. 1998, unpublished.
\bibitem{moretti}A. Moretti, \textit{Rf Update}, presented at the Mini-Workshop on Cooling, BNL Jan. 1998, unpublished.
\bibitem{dave}D. Neuffer and A. Van Ginneken, \textit{Recent Cooling Simulation Studies},  presented at the Mini-Workshop on Cooling, BNL Jan. 1998, unpublished.
\bibitem{balbekov}V. Balbekov and A. Van Ginneken, \textit{Ring Cooler for Muon Collider},  presented at the Mini-Workshop on Cooling, Fermilab Oct. 1997, unpublished. 
\bibitem{wan}D. Neuffer and W. Wan, \textit{COSY transport for $\mu$ cooling}, presented at the Mini-Workshop on Cooling, Fermilab Oct. 1997, unpublished.  
 \bibitem{ref26}G. Silvestrov,  Proceedings of the Muon Collider Workshop,
February 22, 1993, Los Alamos National
  Laboratory Report LA-UR-93-866 (1993); B. Bayanov, J. Petrov, G. Silvestrov,
 J. MacLachlan, and G.
Nicholls, Nucl. Inst. and Meth. {\bf 190}, (1981) 9; C. D. Johnson,
Hyperfine Interactions, {\bf 44} (1988) 21; M. D. Church and J. P. Marriner,
Annu. Rev. Nucl. Sci. {\bf 43} (1993) 253.
 \bibitem{ref27} G. Silvestrov, {\it Lithium Lenses for Muon Colliders}, 
Proceedings of
the 9th Advanced ICFA Beam Dynamics Workshop, Ed. J. C. 
Gallardo, AIP Press, Conference Proceedings 372 (1996).
\bibitem{ref28}F. Mills and C. Johnstone, presentation at the 4th Int. Conference on Physics Potential and Development of $\mu-\mu$ Colliders, San Francisco, CA, Dec. 1997.
 \bibitem{ref29}D. Summers, presentation at the 9th Advanced ICFA Beam Dynamics
Workshop, Montauk 1995, unpublished.
\bibitem{ref30}G.  Morgan, presentation at the 9th Advanced ICFA Beam Dynamics
Workshop, Montauk 1995, unpublished.
 \bibitem{ref31}D. Neuffer, {\it Acceleration to Collisions for the
$\mu^+\mu^-$ Collider}, Proceedings of
the 9th Advanced ICFA Beam Dynamics Workshop, Ed. J. C. 
Gallardo, AIP Press, Conference Proceedings 372 (1996).
 \bibitem{ref32}S.Y. Lee, K.-Y. Ng and D. Trbojevic, FNAL Report FN595 (1992);
Phys. Rev. {\bf E48}, (1993) 3040; D. Trbojevic, et al., {\it Design of the Muon Collider
Isochronous Storage Ring Lattice}, {\it Micro-Bunches Workshop}, BNL Oct.
(1995), AIP Press, Conference Proceedings 367 (1996).
\bibitem{chromatic}K. L. Brown and J. Spencer, SLAC-PUB-2678 (1981) presented at the Particle
Accelerator Conf., Washington, (1981) and
K.L. Brown, SLAC-PUB-4811 (1988), Proc. Capri Workshop, June 1988 and
J.J. Murray, K. L. Brown and T.H. Fieguth, Particle  Accelerator
Conf., Washington, 1987; Bruce Dunham and Olivier Napoly, {\it FFADA, Final Focus.
Automatic Design and Analysis}, CERN Report CLIC Note 222, (1994); Olivier
Napoly, {it CLIC Final Focus System: Upgraded Version with Increased Bandwidth
and Error Analysis}, CERN Report CLIC Note 227, (1994).
\bibitem{ref33}A. Garren, C. Johnstone, \textit{Lattice Design for a 100 GeV Muon Collider}, presentation at the 4th Int. Conference on Physics Potential and Development of $\mu-\mu$ Colliders, San Francisco, CA, Dec. 1997; A. Garren and C. Johnstone, \textit{Progress on a Lattice for a 2 TeV Muon Collider}, submitted to the Proceedings of the PAC97, Vancouver, Canada, June 1997. 
\bibitem{ref33a}D. Trbojevic and  K.-Y. Ng, submitted to
Proc. of the 4th Int. Conf. on Physics Potential and Development
at mu+mu- Colliders, San Francisco, CA, December 10-12, 1997
\bibitem{ref34}A. Drozhdin, C. Johnstone and  N. Mokhov, \textit{Muon Collider Beam Collimation System}, unpublished.
\bibitem{furman}M. Furman, \textit{The Classical Beam-Beam Interaction for the Muon Collider: A First Look}, BF-19/CBP-Note-169/LBL-38563, April 1996. 
\bibitem{chen}P. Chen, \textit{Beam-Beam interaction at $\mu^+\mu^-$ Colliders}, in Proceedings of the Symposium on 
Physics Potential \& Development of $\mu^+\mu^-$ 
Colliders, Nucl. Phys B (Proc. Suppl.) {\bf 51A} (1996);  

\bibitem{ref35}W.-H. Cheng, A. M. Sessler and J. Wurtele, \textit{Studies of Collective Instabilities in Muon Collider Rings},  Proceedings of
the 9th Advanced ICFA Beam Dynamics Workshop, Ed. J. C. 
Gallardo, AIP Press, Conference Proceedings 372 (1996).
  \bibitem{ref36}A. Chao, {\it Physics of Collective Beam Instabilities in
High Energy Accelerators}, John Wiley \& Sons, Inc, New York (1993).
  \bibitem{ref38}K.Y. Ng, {\em Beam Stability Issues in a Quasi-Isochronous 
Muon Collider}, Proceedings of
the 9th Advanced ICFA Beam Dynamics Workshop, Ed. J. C. 
Gallardo, AIP Press, Conference Proceedings 372 (1996).
\bibitem{ref39}I. Stumer et al., \textit{Study of Detector Backgrounds in a $\mu^+\mu^-$ Collider}, Proceedings of the 1996 DPF/DPB Summer Study on High-Energy Physics, Snowmass'96.
\bibitem{ref40}\textit{ Geant Manual}, Cern Program Library V. 3.21, Geneva,
Switzerland, 1993.
\bibitem{ref41}B. King, presentation at the Muon Collider Mini-Workshop: Lattice and Background, UCLA, Feb. 1997 and private communication.  
\bibitem{mokhovrad}N. V. Mokhov and A. Van Ginneken, \textit{Muon Collider Neutrino Radiation}, presentation at the Muon Collider Collaboration Meeting, Orcas Is., Washington (1997); C.J. Johnstone and N.V. Mokhov, \textit{Shielding the Muon Collider Interaction Region}, presented at the PAC97, Vancouver, Canada, 1997.
\bibitem{ref42} G. W. Foster and N. V. Mokhov, {\it Backgrounds and Detector
Performance at 2 + 2 TeV $\mu^+\mu^-$ Collider}, {\it Physics Potential \&
Development of $\mu^+\mu^-$ Colliders
2$^{nd}$ Workshop}, Sausalito, CA, Ed. D. Cline, AIP Press, Woodbury, New York,
pp. 178  (1995).
\end{thebibliography}
